\documentclass[prl,superscriptaddress,twocolumn]{revtex4-1}
\usepackage{times,amsmath,upgreek}
\usepackage{epsfig}
\usepackage{graphicx}
\usepackage{dcolumn}
\usepackage{bm}
\usepackage{tabularx}
\usepackage{hyperref}
\hypersetup{citecolor=black, filecolor= black, linkcolor= black, urlcolor= black}
\begin{document}
\preprint{version 0}

\title{Evolutionary Metadynamics: a Novel Method to Predict Crystal Structures}

\author{Qiang Zhu}
\email{qiang.zhu@stonybrook.edu}
\affiliation{Department of Geosciences, Department of Physics and Astronomy, and New York Center for Computational Sciences, Stony Brook University, Stony Brook, New York 11794, USA}
\author{Artem R. Oganov}
\affiliation{Department of Geosciences, Department of Physics and Astronomy, and New York Center for Computational Sciences, Stony Brook University, Stony Brook, New York 11794, USA}
\affiliation{Geology Department, Moscow State University, 119992, Moscow, Russia}
\author{Andriy O. Lyakhov}
\affiliation{Department of Geosciences, Department of Physics and Astronomy, and New York Center for Computational Sciences, Stony Brook University, Stony Brook, New York 11794, USA}

\begin{abstract}
{A novel method for crystal structure prediction, based on metadynamics and evolutionary algorithms, is presented here. This technique can be used to produce efficiently both the ground state and metastable states easily reachable from a reasonable initial structure. We use the cell shape as collective variable and evolutionary variation operators developed in the context of the USPEX method [Oganov, Glass, \textit{J. Chem. Phys.}, 2006, \textbf{124}, 244704; Lyakhov \textit{et al., Comp. Phys. Comm.}, 2010, \textbf{181}, 1623; Oganov \textit{et al., Acc. Chem. Res.}, 2011, \textbf{44}, 227] to equilibrate the system as a function of the collective variables. We illustrate how this approach helps one to find stable and metastable states for Al$_2$SiO$_5$, SiO$_2$, MgSiO$_3$, and carbon. Apart from predicting crystal structures, the new method can also provide insight into mechanisms of phase transitions.}
\end{abstract}
\maketitle
\section{Introduction}

Crystal structure prediction (CSP) has long been a major fundamental challenge in physical sciences \cite{Maddox-Nature-1988}. As the stable structure corresponds to the global minimum of the free energy surface (FES), several global optimization algorithms have been devised and applied \cite{Pannetier-Nature-1990, Martonak-PRL-2003, Wales-1997-JPC, Goedecker-JCP-2004, Oganov-JCP-2006, Freeman-JMC-1993}. Due to these efforts, in many cases it is now possible to predict the stable structure of a given inorganic compound at arbitrary external pressure. In addition, there is progress towards optimizing not only free energy, but also other properties (e.g., hardness \cite{Lyakhov-PRB-2011}, density \cite{Zhu-PRB-2011}, etc).

In general, there are two types of strategies for predicting crystal structures. One is to directly scan the whole energy landscape, and find the most stable crystal structures using random sampling \cite{Freeman-JMC-1993} or generally more efficient evolutionary algorithms \cite{Oganov-JCP-2006}. Alternatively, one could also use some known structures as the starting point, and predict the new structures by crossing the energy barriers, until the lowest energy structure is found \cite{Pannetier-Nature-1990, Martonak-PRL-2003, Wales-1997-JPC, Goedecker-JCP-2004}. The latter group of methods can be described as neighborhood search methods and some of them - in particular, metadynamics \cite{Laio-PNAS-2002, Martonak-PRL-2003}, can be very efficient, but rely on availability of a good starting structure.

Metadynamics explores the properties of the multidimensional FES of complex many-body systems by means of a coarse-grained non-Markovian dynamics in the space defined by a few collective coordinates. By introducing a history-dependent potential term, it fills the minima in the FES and allows efficient exploration of the FES as a function of the collective coordinates \cite{Laio-PNAS-2002}. The technique is usually applied as an extension of the molecular dynamics (MD) simulation technique. Martonak \emph{et al.} used the edges of the simulation cell as collective variables for the study of pressure-induced structural transformations \cite{Martonak-PRL-2003}. The method proved to be much more powerful in predicting crystal structure transformations \cite{Oganov-Nature-2005, Martonak-NM-2006, Raiteri-Angew-2005}, compared with the normal MD approach. However, at each metastep it uses MD for equilibrating the system and MD is not always an efficient method for equilibration, which leads to trapping in metastable states and often amorphization rather than transition to a stable crystal structure. This motivated us to develop an alternative strategy.

In this paper, we present a method combining the features of both strategies for predicting crystal structures, basically a metadynamics-like method driven not by local MD sampling, but by efficient global optimization moves \cite{Oganov-JCP-2006, Lyakhov-CPC-2010}. Following Martonak \emph{et al.} \cite{Martonak-PRL-2003}, we adopt the cell edges as collective variables, and equilibrate the system at each value of the collective variables using moves inspired by the evolutionary variation operators \cite{Lyakhov-CPC-2010}, rather than previously used MD simulation. We find that this approach is very efficient for predicting stable and low-energy metastable states and avoids amorphization.

\section{Methodology}
MD method is widely used to study physical processes in liquids and solids \cite{Parrinello-PRL-1980}, but has difficulties in escaping from local energy minima and crossing high energy barriers \cite{Voter-JCP-1997}. Metadynamics \cite{Laio-PNAS-2002} is an ingenious way to solve this problem and accelerate the activated processes involving barrier crossing, including first-order reconstructive transitions. In metadynamics, one has to specify collective variables that adequately describe the evolution of the system, and then perturb the FES defined in this reduced space of collective variables. Martonak \emph{et al.} used the cell box \emph{h} (3 $\times$ 3 matrix) as a collective variable to distinguish different structures \cite{Martonak-PRL-2003}. For a given system with volume \emph{V} under external pressure \emph{P}, the derivative of the free energy \emph{G} with respect to \emph{h} is
\begin{equation} \label{eq:Gibbs1}
-\frac{\partial G}{\partial h_{ij}} = V[h^{-1}(P-p)]_{ji}
\end{equation}
where \emph{p} is the internal pressure tensor calculated for each given geometry.

The collective variable evolves with a stepping parameter $\delta$\emph{h},
\begin{equation} \label{eq:tensor}
h(t+1) = h(t) + \delta{h}\frac{f(t)}{|f(t)|}
\end{equation}
here the driving force $f = -\frac{\partial G}{\partial h}$ is from a history-dependent Gibbs potential $G(t)$ where a Gaussian has been added to $G^h$ at every point $h(t')$ already visited in order to discourage it from being visited again,
\begin{equation} \label{eq:Gibbs}
G(t) = G^h + \sum We^{-\frac{|h-h(t')|^2}{2\delta{h^2}}}
\end{equation}

As the simulation proceeds, the history-dependent term fills the initial well of the FES, and concurrently the collective variables (the cell) undergo a sequence of changes, at each of which the atoms are re-equilibrated. At some critical cell distortion atoms rearrange dramatically, yielding a new crystal structure.

By adding a history-dependent Gaussian potential, metadynamics efficiently reaches and crosses the transition state, thus solving an intrinsic problem of MD simulations. However, it still relies on the ability of MD to equilibrate the system at each value of collective variables. Metadynamics encounters two general problems. First, MD is not a very efficient technique for equilibration involving crossing energy barriers, especially at low temperatures. Second, metadynamics reduces the full energy landscape to a low-dimensional projection, which is not always adequate.

For a crystal with \emph{N} atoms in the unit cell, the number of degrees of freedom is 3\emph{N}+3. Metadynamics assumes that the system can be described by 6 collective variables representing the box \emph{h}. The remaining 3\emph{N}-3 dimensions describe atomic positions, and we need to determine the global energy minimum with respect to these dimensions at given \emph{h}. This can be done using MD (as in standard metadynamics), but as discussed above, this encounters certain problems. Here we do it using global optimization techniques based on evolutionary algorithms \cite{Oganov-JCP-2006, Lyakhov-CPC-2010, Oganov-ACR-2011}. These algorithms use several variation operators - i.e. recipes for obtaining new structures from the previously sampled ones - for sampling the energy landscape. Here we use the softmutation operator \cite{Lyakhov-CPC-2010}, which connects global optimization with lattice dynamics and group theory. Indeed, the 3\emph{N}-3 variables describing atomic positions can be transformed into a set of normal modes that possess valuable properties. If a structure is not dynamically stable, a more stable structure is obtained by following the eigenvector of the softest vibrational mode. For structures without soft modes, there is a statistically valid Bell-Evans-Polanyi principle that states that low-energy structures are usually connected by low activation barriers \cite{Jensen-1999}. Low barriers are, in turn, usually related to the direction of the lowest curvature of the FES - or eigenvectors of the softest vibrational mode. This is the principle behind the minima hopping method \cite{Goedecker-JCP-2004} and explains the efficiency of the softmutation operator introduced in the evolutionary algorithm USPEX \cite{Lyakhov-CPC-2010}. To calculate the vibrational modes, we construct the dynamical matrix from bond hardness coefficients.
\begin{equation} \label{eq:dynamical matrix}
\begin{aligned}
D_{\alpha\beta}(a,b) = &\sum_m \Big(\frac {\partial ^2}{{\partial \alpha_a^0}{\partial \beta_b^m}} \frac{1}{2} \sum_{i,j,l,n} H_{i,j}^{l,n}(r_{i,j}^{l,n}-{r_{0}}_{i,j}^{l,n})^2\Big)
\end{aligned}
\end{equation}
Here coefficients $\alpha$, $\beta$ denote coordinates $(x,y,z)$; coefficients $a, b, i, j$ describe the atom in the unit cell; coefficients $l, m, n$ describe the unit cell number; $r_{i,j}^{l,n}$ is the distance between atom $i$ in the unit cell $l$ and atom $j$ in the unit cell $n$, while ${r_{0}}_{i,j}^{l,n}$ is corresponding bond distance, and $H_{i,j}^{l,n}$ is bond hardness coefficient computed from bond distances, covalent radii and electronegativities of the atoms \cite{Lyakhov-CPC-2010, Li-PRL-2008}.

To perform softmutation, we move the atoms along the eigenvector of the softest calculated mode. One structure can be softmutated many times using different non-degenerate modes and displacements. By properties of normal modes, the original and softmutated structures are linked by group-subgroup symmetry relations, but structure relaxation may result in a symmetry increase. In this case one observes a structural transition with a common subgroup. The magnitude of the displacement ($d_{\rm max}$) along the mode eigenvector is an input parameter: with relatively small $d_{\rm max}$ and displacements represented by a random linear mixture of all mode eigenvectors, we obtain a method similar to MD-metadynamics in its ability to cross energy barriers and equilibrate the system. With large $d_{\rm max}$ along the softest mode eigenvectors, we obtain the softmutation operator \cite{Lyakhov-CPC-2010}, capable of efficiently finding the global energy minimum. When needed, other evolutionary variation operators can be used - such as permutation (swaps of atomic identities) or heredity (combination of pieces of two parent structures).

\begin{figure}
\epsfig{file=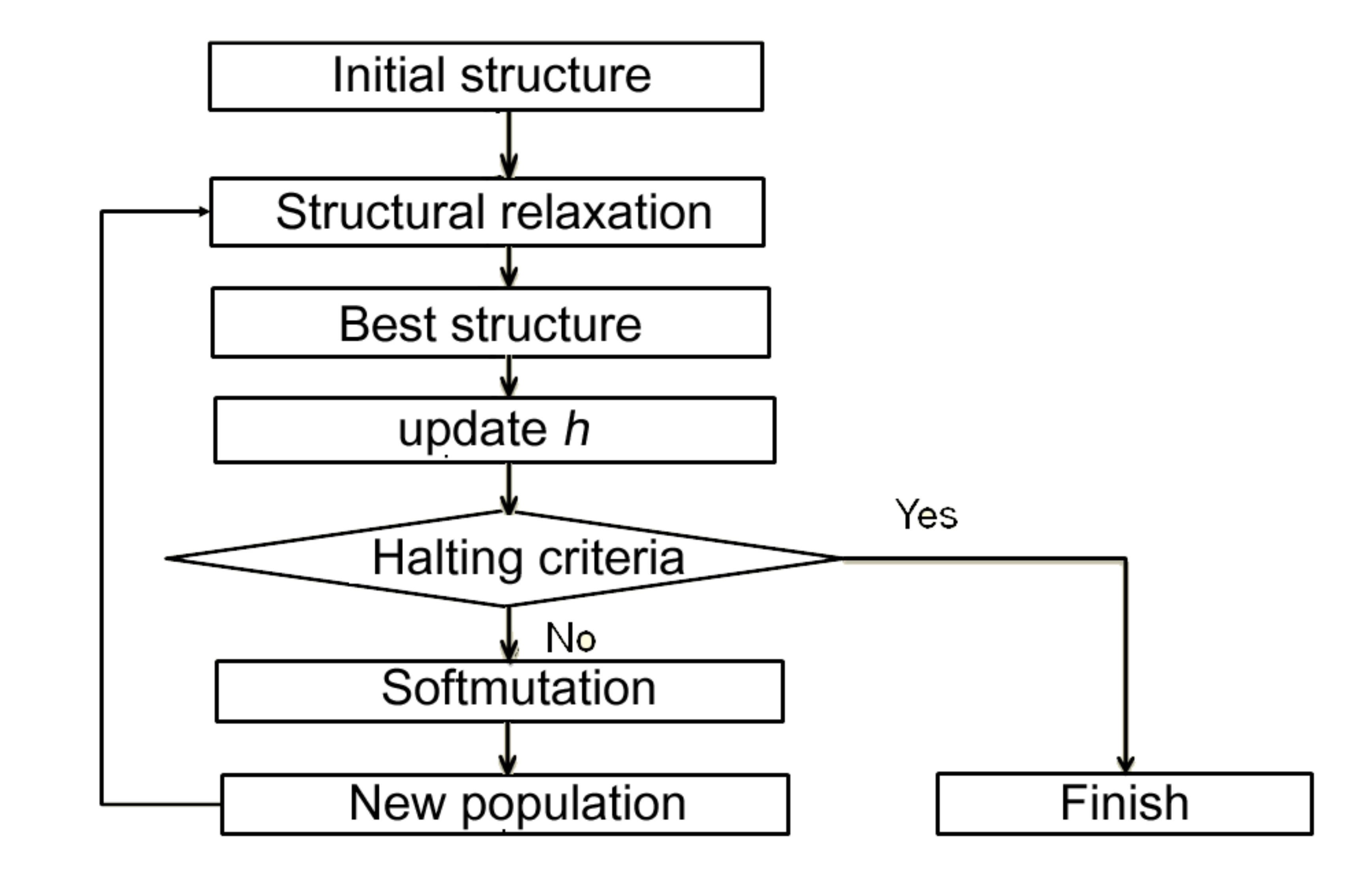, width=0.35\textwidth}
\caption{\label{flowchart} Illustration of the evolutionary metadynamics algorithm.}
\end{figure}

Our algorithm as shown in Fig. \ref{flowchart} is in many ways similar to the original version by Martonak \emph{et al.} \cite{Martonak-ZFK-2005}. We start from one known initial structure at a given external pressure \emph{P}, and then compute its vibrational modes, which are used to produce typically 20-40 softmutated structures. These are relaxed at constant \emph{h}, the lowest-enthalpy structure is selected and its pressure tensor \emph{p} is computed. Its box \emph{h} is then updated according to Eq. (\ref{eq:tensor}), and a new generation of softmutated structures are produced and relaxed in the fixed cell \emph{h}. Repeated for a number of generations, this computational scheme leads to a series of structural transitions and is stopped when the maximum number of generations is reached. Due to the presence of a population of structures and a selection step, this algorithm is evolutionary, unlike original metadynamics \cite{Martonak-ZFK-2005}.

In addition to $d_{\rm max}$, there are two other important parameters - the Gaussian height ($W$) and Gaussian width ($\delta{h}$), for choosing which Martonak \emph{et al}. \cite{Martonak-ZFK-2005} proposed a recipe based on the curvature of the FES at the minimum. According to our experience, a good starting point is to set $\delta{h}$ as one tenth of the minimum lattice vector, and $W$ as around 8,000 $\delta{h}^2$ (with the unit of kbar$\cdot$\AA$^3$).

Note that Eq. (\ref{eq:tensor}) is not invariant with respect to the choice of the unit cell (or supercell) and works best for cells close to cubic shape. To remedy this, we developed a more sophisticated equation based on elasticity theory,
\begin{equation} \label{eq:newtensor}
h_{im}(t+1) = h_{im}(t)+\frac{\delta h}{|f|V^{1/3}}S_{ijkl}f_{kl}h_{jm}(t)
\end{equation}
Here we use the elastic compliance tensor ($S$) corresponding to an elastically isotropic medium with Poisson ratio 0.26, which corresponds to the border between brittle and ductile materials \cite{Pugh-1954} and is a good average value to describe both metals and insulators. The choice of the Poisson ratio does not significantly affect the results; the main effect of Eq. (\ref{eq:newtensor}) is to make the results independent of the simulation cell shape. The Young's modulus is chosen in such a way that for a cubic cell under uniform stress the original formula Eq. (\ref{eq:tensor}) is reproduced. We have implemented and tested the formalism based on Eq. (\ref{eq:newtensor}) and found it to work extremely well, the efficiency improves compared to Eq. (\ref{eq:tensor}) (results not shown here). 

\begin{figure}
\epsfig{file=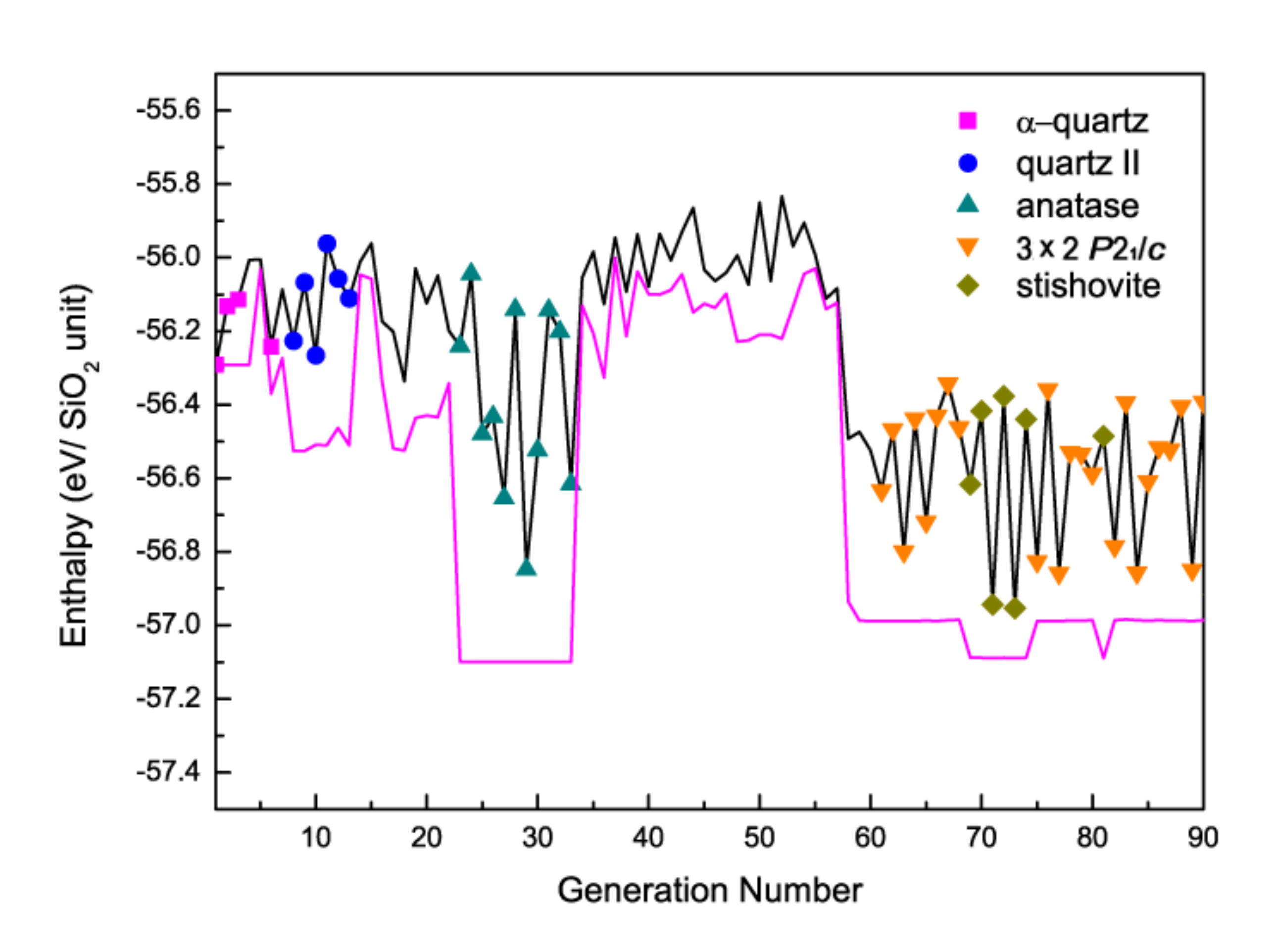, width=0.35\textwidth}
\caption{\label{SiO2}Enthalpy evolution during the compression on 72-atom supercell of $\alpha$-quartz (SiO$_2$) at 10 GPa (black line: enthalpies for best structures with constant \emph{h}; magenta line: enthalpies for best structures after full relaxation). In this calculation, we set Gaussian parameters $W$=10000 kbar$\cdot$\AA$^3$, $\delta{h}$=1.0 \AA, and $d_{\rm max}$=3.0 \AA. Each generation contains 40 structures. Structures were relaxed using the GULP code \cite{GULP-2003} with the BKS potential \cite{BKS}. We first observed the transition from $\alpha$-quartz (space group \emph{P}3$_1$21) to quartz II (\emph{C}2) in the 9th generation, and then quartz II amorphized until it transformed into the anatase structure (\emph{I}4$_1$/\emph{amd}) in the 22th generation. Anatase amorphized again, and evolved into the 3 $\times$ 2 \emph{P}2$_1$/\emph{c} structure in the 58th generation, and then transformed into stishovite (\emph{P}4$_2$/\emph{mnm}) at the 69th generation.}
\end{figure}

\begin{figure*}
\epsfig{file=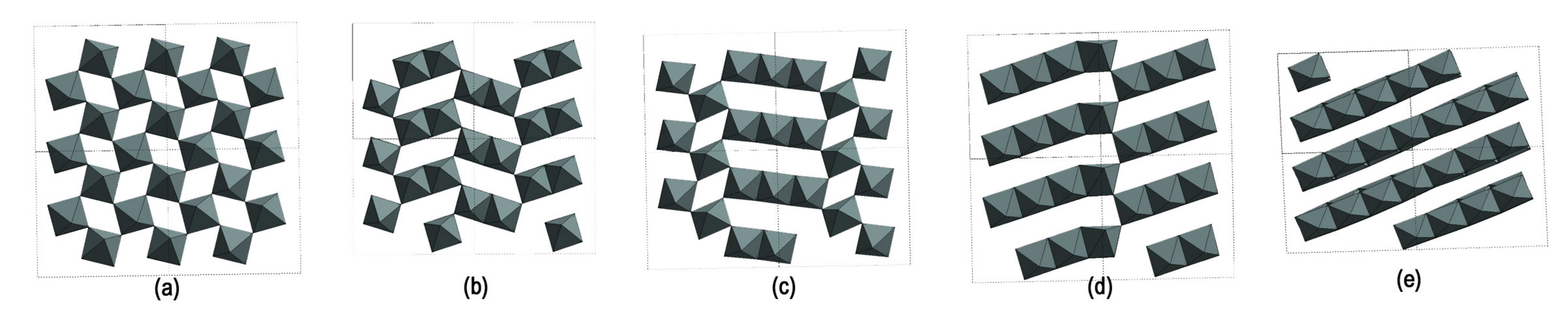, width=0.85\textwidth}
\caption{\label{mgsio3}Phases observed in evolutionary metadynamics starting from a 160-atom supercell of perovskite (MgSiO$_3$) at 250 GPa. (a) perovskite (space group \emph{Pbnm}); (b) 2 $\times$ 2 phase (\emph{Pbnm}); (c) 3 $\times$ 1 phase (\emph{P}2$_1$/\emph{m}); (d) 4 $\times$ 4 phase (\emph{Pm}); (e) post-perovskite (\emph{Cmcm}). Only Si octahedra are shown (Mg atoms are omitted for clarity). In this calculation, we set Gaussian parameters $W$=7000 kbar$\cdot$\AA$^3$, $\delta{h}$=0.8 \AA, and $d_{\rm max}$=3.0 \AA.  Each generation contains 40 structures. Structures were relaxed using the GULP code \cite{GULP-2003} with a partially ionic Buckingham potential \cite{Oganov-PEPI-2000}. Our calculation found the transition from perovskite to post-perovskite, and structures b, c, d were also observed in the simulation. Plane sliding with the formation of stacking faults is a possible pathway for this phase transition, in accordance with the previous metadynamics study \cite{Martonak-NM-2006}.}
\end{figure*}

Below we discuss results obtained with Eq. (\ref{eq:tensor}) and Eq. (\ref{eq:Gibbs}), \emph{i.e.} the original and simplest formulation of metadynamics \cite{Martonak-ZFK-2005} with isotropic Gaussians. In reality, the FES minima are anisotropic - the local FES curvature is lower for shear and higher for compressional distortions of the cell. It was found that the "isotropic" formalism based on Eq. (\ref{eq:Gibbs}) coupled with MD equilibration is incapable of predicting structural transformations of silica and gets stuck in amorphization \cite{Martonak-NM-2006}. This problem was remedied by the anisotropic extension of Eq. (\ref{eq:Gibbs}) \cite{Martonak-NM-2006}. Here we find that this increase of complexity can be avoided and the same structural transformations are easily predicted with the isotropic formulation of metadynamics, if softmutation is used instead of MD for equilibration (see Fig. \ref{SiO2}). We also checked that the approach presented here correctly reproduces the previous results on pressure-induced transitions in MgSiO$_3$ \cite{Oganov-Nature-2005} (see Fig. \ref{mgsio3}). After these successful tests, we applied it to two challenging and important problems, namely the pressure-induced transformations of elemental carbon and Al$_2$SiO$_5$. Modified formalism based on Eq. (\ref{eq:newtensor}) gives very similar results (not shown here), but is invariant to the choice of supercell and more efficient.

\section{Results and Discussion}
\subsection{Finding stable and metastable energy minima: example of Al$_2$SiO$_5$}
The phase diagram of Al$_2$SiO$_5$ is important for Earth sciences and has attracted great interest. Three known Al$_2$SiO$_5$ polymorphs (kyanite, andalusite, sillimanite) are common minerals in the Earth's crust and upper mantle. In all these structures, Si atoms are tetrahedrally coordinated, while half of the Al atoms are octahedrally coordinated. The Al octahedra form the -Al-Al- chains, and the remaining Al and Si alternate in neighboring chains (-Si-Al-). The coordination of Al in the -Si-Al- chains is either tetrahedral (sillimanite), fivefold (andalusite), or octahedral (kyanite) \cite{Oganov-acta-2001}. It has to be noted that the prediction of these structures is an extremely challenging test for any global optimization method (These complex structures have low symmetry and relatively large primitive cells with 32 atoms. To illustrate the difficulty of finding the ground state, we generated 10,000 random structures and relaxed them at 10 GPa, and found that none of these structures correspond to the stable phase, kyanite). Energy barriers between these structures are very high and these phases can exist metastably, and even coexist, in nature for millions of years - making direct MD sampling (which covers timescales up to $\upmu$s) of these structural transitions clearly impossible. Impressively, an evolutionary metadynamics simulating starting from the low-pressure polymorph, andalusite, has successfully found the other two structures.

\begin{figure}
\epsfig{file=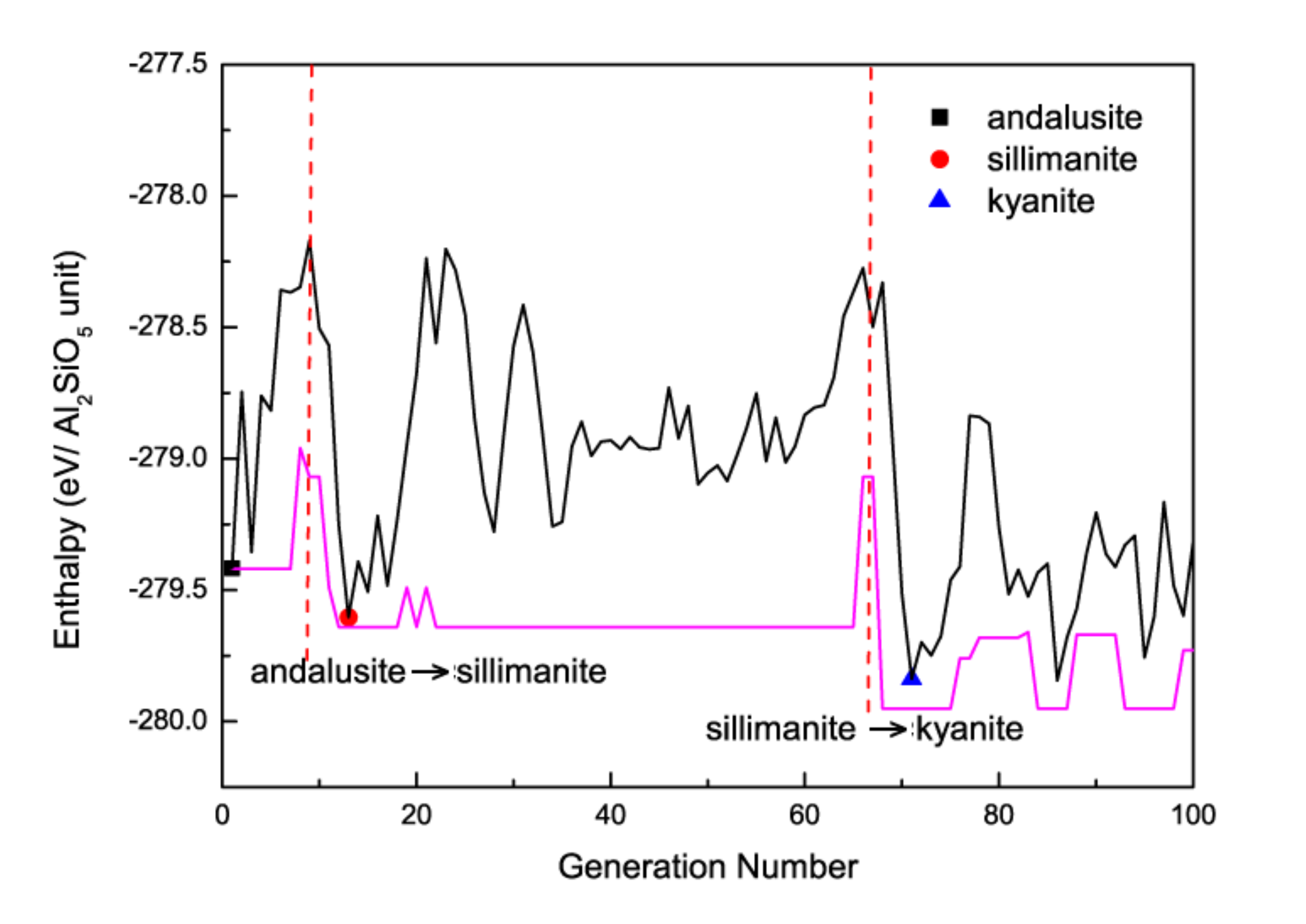, width=0.35\textwidth}
\caption{\label{al}Enthalpy evolution during the compression on andalusite (Al$_2$SiO$_5$) at 10 GPa (black line: enthalpies for best structures with constant \emph{h}; magenta line: enthalpies for best structures after full relaxation).}
\end{figure}

\begin{figure*}
\epsfig{file=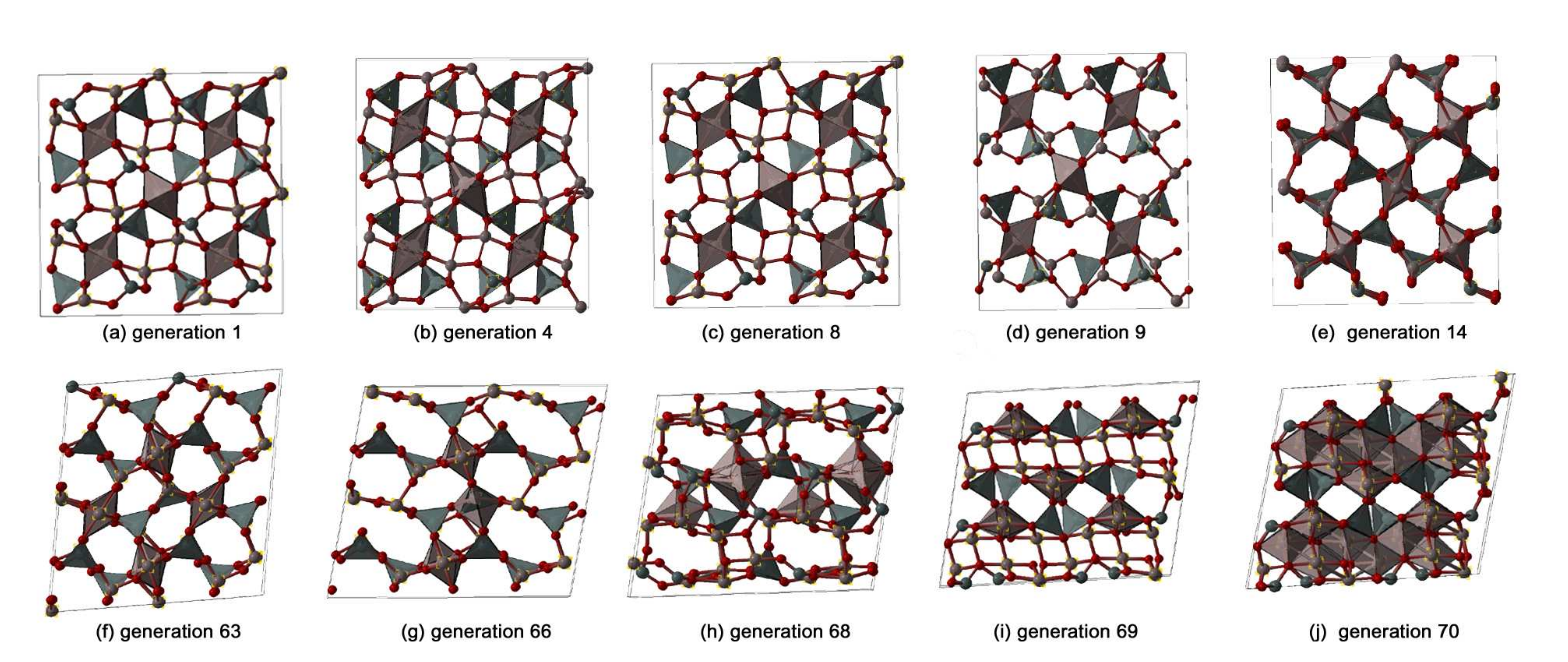, width=0.85\textwidth}
\caption{\label{al2sio5}Phases observed during compressing andalusite (Al$_2$SiO$_5$) at 10 GPa. (a) generation 1 (andalusite); (b) generation 4; (c) generation 8; (d) generation 9 (sillimanite); (e) generation 14; (f) generation 63; (g) generation 66; (h) generation 68; (i) generation 69; (j) generation 70 (kyanite).}
\end{figure*}

These simulations were carried out by using a classical potential \cite{Urusov-1998} and the GULP code \cite{GULP-2003}. We started the calculation with $d_{\rm max}$=3.0 \AA, $W$=2500 kbar$\cdot$\AA$^3$ and $\delta{h}$=0.4 \AA.  Each generation contains 30 structures. Starting from andalusite (32 atoms in the supercell), as shown in Fig. \ref{al}, in the 8th generation we observed breaking of an interchain Al-O bond in AlO$_5$ polyhedra, and the structure transformed into sillimanite containing AlO$_4$ tetrahedra. Sillimanite survived until the 68th generation, when the inter-chain Al-O bonds formed again, increasing coordination of Al from fourfold to sixfold and eventually creating kyanite phase with all Al atoms in the AlO$_6$ octahedra. The whole picture, as shown in Fig. \ref{al2sio5}, proves that our method was easily able to predict the transitions: andalusite $\rightarrow$ sillimanite $\rightarrow$ kyanite. As a bonus in addition to finding the global minimum structure, the simulation unravels a very non-trivial relationship between the structures (e.g. sillimanite-kyanite) and a crystallographic model of their transformations. For a reconstructive phase transition, one should expect a nucleation-and-growth mechanism, rather than a concerted crystallographic mechanism. However, such a concerted mechanism provides not only a useful simplified view of the real transition, but also input for mean-field theories of phase transformations, and for techniques (such as the Transition Path Sampling(TPS) \cite{Boulfelfel-PNAS-2011}) that are capable of simulating nucleation and growth phenomena but require a reasonable initial mechanism. 

\subsection{Searching for low-energy metastable phases: elemental carbon}
Carbon is unique in that it adopts a wide range of structures, from superhard insulating (diamond, lonsdaleite, and new allotrope \cite{Mao-Science-2003}) to ultrasoft semi-metallic (graphite, fullerenes) and even superconducting (doped diamond \cite{Ekimov-Nature-2004} and alkali-doped fullerenes \cite{Tanigaki-Nature-1991}). The number of possible metastable phases is infinite. By compressing graphite at room temperature to 15-20 GPa, a metastable transparent superhard phase was phase observed, but its structure remained uncertain \cite{Mao-Science-2003}. Several structural models were proposed \cite{Li-PRL-2009, Umemoto-PRL-2010}. The correct model should have the lowest barrier of formation from graphite at 15-20 GPa. To determine the barrier, it is necessary to study the transition pathways from graphite to all candidate structures. The technique discussed here is capable of finding several (if not all) relevant candidates in one single simulation.

\begin{figure}
\epsfig{file=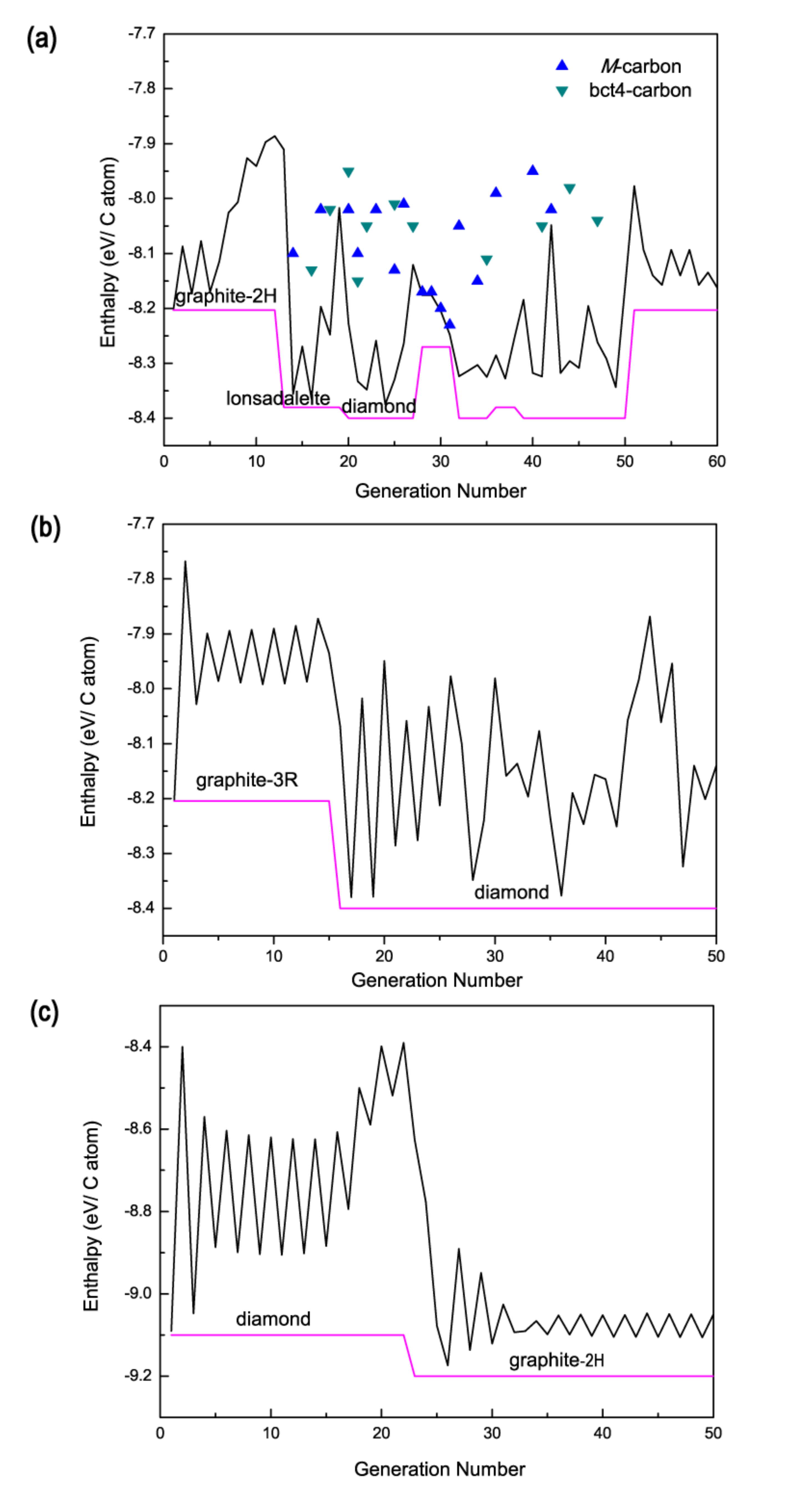, width=0.35\textwidth}
\caption{\label{C-meta}Enthalpy evolution of (a) graphite-2H at 20 GPa; (b) graphite-3R at 20 GPa; (c) diamond at 0 GPa (black line: enthalpies for best structures with constant \emph{h}; magenta line: enthalpies for best structures after full relaxation).}

\end{figure}

\begin{figure*}
\epsfig{file=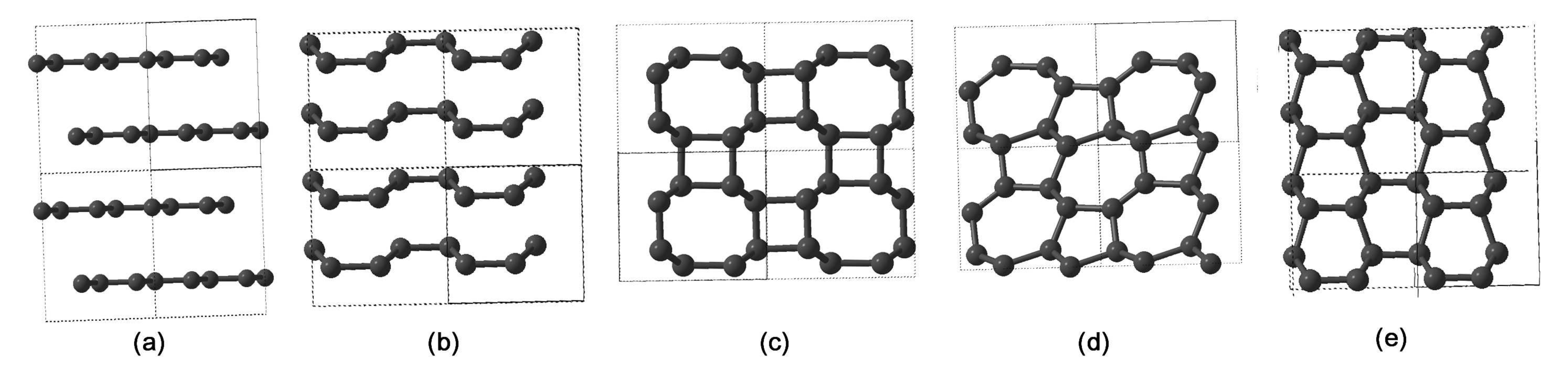, width=0.85\textwidth}
\caption{\label{graphite}Phases observed during compressing graphite-2H at 20 GPa. (a) initial graphite-2H; (b) buckled graphite layer; (c) bct C$_4$ with 4+8 membered rings; (d) \emph{M}-carbon with 5+7 membered rings; (e) lonsdaleite.}
\end{figure*}

We started the simulation at 20 GPa from the graphite-2H structure with 32 atoms per box and using $d_{\rm max}$=2.5 \AA, $W$=7000 kbar$\cdot$\AA$^3$ and $\delta{h}$=0.7 \AA.  Each generation contains 25 structures. Structure relaxations were done using density functional theory (DFT) within the generalized gradient approximation (GGA) \cite{GGA-1996} together with the all-electron projector augmented wave (PAW) \cite{PAW-1994} method as implemented in the VASP code \cite{VASP-1996}.  We used the plane wave kinetic energy cutoff of 520 eV and the Brillouin zone sampling resolution of 2$\pi$ $\times$ 0.08 \AA$^{-1}$, which showed excellent convergences of the energy differences, stress tensors and structural parameters. Fig. \ref{C-meta}a shows the results. In the 13th generation, graphene layers buckled as shown in Fig. \ref{graphite}b, which led to the formation of 3D networks of $sp^3$-hybridized carbon atoms. We observed, in a single simulation, the formation of lonsdaleite with 6-membered rings in the 14th generation (Fig. \ref{graphite}e), \emph{M}-carbon containing 5- and 7-membered rings (Fig. \ref{graphite}d), and bct-C$_4$ structure with 4- and 8-membered rings (Fig. \ref{graphite}c) in the 16th generation. In the 20th generation, global minimum, diamond, was found. Diamond is dominant in most of the following generations, except some occurences of lonsdaleite or \emph{M}-carbon as the lowest-energy structure in a generation. At the 51th generation, the system reverted to graphite. We note that since at each value of \emph{h} a number of structures are obtained by softmutation at each generation, or metastep (but only one structure was found using MD moves in the original metadynamics method \cite{Martonak-PRL-2003}), the structural information is much richer in this version of metadynamics. Many candidate metastable structures can thus be produced.

Both bct-C$_4$ and \emph{M}-carbon have been shown to match the experimental diffraction patterns \cite{Li-PRL-2009, Umemoto-PRL-2010}, and it was unclear which structure has the lower formation barrier. The best method to compute these barriers is TPS, but it requires an initial model of the pathway, which is then evolved using Monte Carlo sampling. Our results provide such an initial pathway both for the graphite to \emph{M}-carbon and graphite to bct-C$_4$ transformations and structural relationships are very clear from Fig. \ref{graphite}. Our TPS simulations \cite{Boulfelfel-PNAS-2011} suggest that \emph{M}-carbon has the lowest barrier for the formation from graphite among all carbon phases, and is thus the likeliest structure for the phase observed in experiment. This example shows how the present metadynamics-based technique can be used in the search for metastable phases.

Starting the calculation at 20 GPa from another polytype, graphite-3R, we again easily found the diamond structure (ground state) and a number of low-energy metastable structures with $sp^3$ hybridization. We also performed a search at \emph{P} = 1 atm, starting from the diamond structure. Graphite-2H (the ground state) and a number of mixed $sp^2$-$sp^3$ structures were easily found. The results of the simulations are shown in Fig. \ref{C-meta}b and c.

\section{Conclusions}
We have developed a novel method crystal structure prediction, based on the ideas of metadynamics and evolutionary global optimization. We use the cell shape (6 degrees of freedom) as the collective variable, sampling which we find and cross the energy barrier, and point out the direction where the phase transition might occur.  At each value of the collective variable \emph{h}, we equilibrate the system using evolutionary variation operators (in the example given here - softmutation), which efficiently explores the remaining 3\emph{N}-3 internal degrees of freedom. The success of tests on carbon and Al$_2$SiO$_5$ proves the power of this method. Going beyond crystal structure prediction, this method also could produce transformation trajectories between phases, and is thus useful for understanding the transition mechanisms. The method marries ideas from the standard MD-metadynamics \cite{Martonak-ZFK-2005} and evolutionary algorithm USPEX \cite{Oganov-JCP-2006}. Like standard metadynamics and unlike USPEX, present technique does require a reasonable starting structure and has the ability to find transition pathways (due to the use of finite displacements along the lowest-frequency mode eigenvectors) and low-energy metastable structures. Yet, unlike MD-metadynamics, our method has a more efficient equilibration and at each metastep produces a set of structures (rather than a single structure), i.e. gives richer chemical information. It avoids amorphization during the simulation - a common problem for MD-metadynamics. For large systems, it can in some cases be more efficient than USPEX, provided a good initial structure. Present technique has a very different philosophy from the USPEX method and in many ways is complementary to it. 

\section{Acknowledgement}
Calculations were performed on the supercomputer of Center for Functional Nanomaterials, Brookhaven National Laboratory, which is supported by the U.S. Department of Energy, Office of Basic Energy Sciences, under contract No. DE-AC02-98CH10086, and on Skif-MSU supercomputer (Moscow State University, Russia) and at Joint Supercomputer Center (Russian Academy of Sciences, Moscow, Russia). This work is funded by DARPA (grant N66001-10-1-4037), National Science Foundation (grant EAR-1114313). We thank Prof. J. D. Gale for help with the GULP code.

\bibliography{biblio} 

\end{document}